# An Insight into the Dynamics of a Dual Active Bridge

Ezekiel Olayiwola Arogunjo and Joseph Olorunfemi Ojo
Laboratory for Electric Machines and Power Electronics
Electrical and Computer Engineering Department,
Tennessee Technological University, Cookeville TN 38505, USA.
eoarogunjo42@tntech.edu and jojo@tntech.edu

*Abstract*—This paper aims at analyzing the effect of the zero dynamics of the Dual Active Bridge Isolated Bidirectional dc-dc converter (DAB) on the dynamics of the complete DAB system. It also explains its influence on controller design for the DAB system. In carrying out these analyses, the state space model of the DAB, as well as the first harmonic approximation (FHA) of the model are derived. The ZVS and the stability analysis of the system are undertaken based on the FHA model of the system. The system is shown to be stable for a constant output voltage operation for the entire power range while it is unstable for the constant power load (CPL) operation for load demands close to the system maximum power. It is also shown that the transformer winding currents are part of the zero dynamic states and are always stable regardless of the operating conditions of the system.

*Keywords—DAB, zero dynamics, FHA modeling, stability analysis.*

I. INTRODUCTION

The dual active bridge isolated bidirectional dc-dc converter (DAB) was first mentioned in 1988[1]. It consists of two single-phase converters connected through a high frequency transformer for galvanic isolation and flexible voltage applications. It is a buck and boost converter and allows the control and bidirectional flow of power. One of the merits of the DAB is its ease of implementation of the soft-switching techniques[2]. This enables the employment of high frequency operation which allows for a significant reduction in sizes of electronic component of the converter, thereby improving its power density. In the analysis of the converter dynamics and controller design of the DAB, the reduced-order and full-order models are usually employed [3]. In the full-order model the complete model equations of the DAB are employed in analyzing the dynamics of the DAB. For the reduced order model, the transformer currents which are purely ac are neglected. The negligible effect of the neglect of the transformer current is possible because the current has a much faster dynamics than the other states, and as will be shown in this paper, the currents are part of the stable zero dynamics of the system. According to [4], zero dynamics are the worst case of the internal dynamics of a system achieved by constraining the non-zero dynamic states of the system to zero.

The performance of the controller designed for a dual active bridge converter greatly depends on the accuracy of the converter's model. Qin et al studied the dynamics of the DAB using its full order model. The authors however noted that the transformer current of the converter makes continuous-time modeling difficult. Hence, average modeling techniques are deployed to derive the input-output transfer functions of the system for its controller design [3]. Wang et al. presented a reduced order average model of the DAB but noted that using the first order harmonic approximation of the model may not be completely accurate for determining the behavior of the system [5]. Bai et al. presented a reduced order model of a DAB by neglecting the transformer current in the full order continuous-time system. From results shown, the dynamic response of the system was close to the response from the simulated full order continuous-time model under steady-state and transient conditions [6]. This paper therefore aims to use the phenomenon of zero dynamics to show the reason for the negligible effect of the neglect of the transformer current dynamics on the overall response of the system and controller design accuracy as well as the controlled system stability.

This paper is organized as follows:

In Section 1, a general introduction of dual active bridge and zero dynamics is done. In Section 2, a description of the system under consideration is presented. The physical system model and FHA of the continuous-time model is carried out. The steady state analysis, ZVS and stability analysis based on the FHA of the model are presented. In Section 3, the controllability, observability and zero dynamics of the system are addressed.

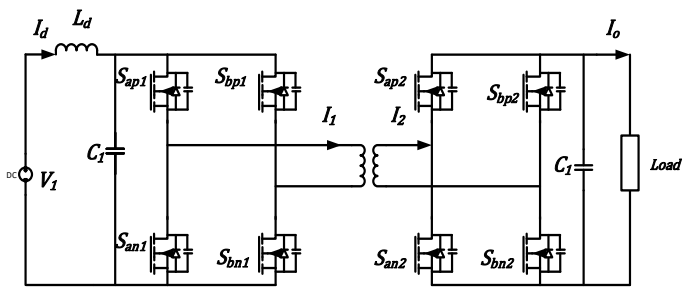

Figure 1. Topology of a dual active bridge dc-dc converter

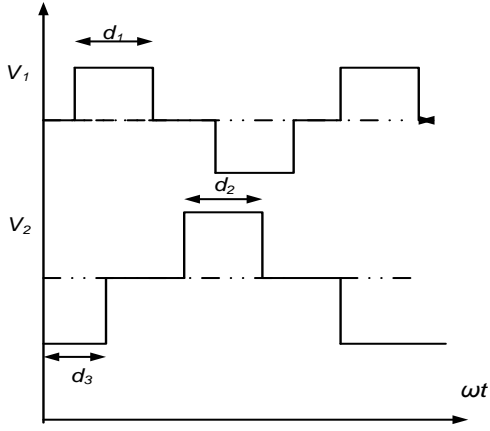

Figure 2: DAB phase shift and duty cycle description

## II. DESCRIPTION, MODELING AND STEAD-STATE ANALYSIS OF SYSTEM.

### A. System Desscription

The DAB in Figure 1 consists of two full active bridges connected by a high frequency transformer with a turns ratio $n$. A diode-bridge-rectified ac voltage $V_1$ is applied to the primary bridge through a LC filter of inductor $L_d$ and capacitor $C_1$. The output of the secondary bridge is connected through a capacitor $C_2$ to a current source (representing the load current) with current $I_o$. The DAB is operated by triple phase shift (TPS) control method.

As shown in Figure 2, $d_1$ and $d_2$ represent respectively, the primary and secondary voltages pulse widths, while the $d_3$ represents the outer phase shift between the primary and secondary bridges. The phase shift angle between the primary and secondary voltages which has more significance in practical terms is defined in terms of $d_3$ as:

$$\delta = \left(d_3 + \frac{(d_1 - d_2)}{2}\right) \quad (1)$$

### B. Mathematical Model of DAB.

The dynamic equation of the DAB is derived by first writing out the KVL equations for the dc and ac circuit of the system in Figure 1. The dynamic equations of the system are as written below:

$$pI_d = \frac{V_1}{L_d} - \frac{V_{c1}}{L_d} - \frac{r}{L_d} I_d \quad (2)$$

$$pV_{c1} = \frac{I_d}{C_1} - \frac{s_1 I_1}{C_1} \quad (3)$$

$$pI_1 = \frac{L_1 R_1}{h} I_1 + \frac{L_m R_2}{h} I_2 - \frac{L_2}{h} s_1 V_{c1} + \frac{L_m}{h} s_2 V_{c2} \quad (4)$$

$$pI_2 = \frac{L_m R_1}{h} I_1 + \frac{L_1 R_2}{h} I_2 - \frac{L_m}{h} s_1 V_{c1} + \frac{L_1}{h} s_2 V_{c2} \quad (5)$$

$$pV_{c2} = \frac{n s_2 I_2}{C_2} - \frac{I_o}{C_2} \quad (6)$$

$$y_1 = V_{c1}, \quad y_2 = V_{c2}$$
$$x = [I_d \quad V_{c1} \quad I_1 \quad I_2 \quad V_{c2}], u = [s_1 \quad s_2]$$

Where $V_{c1}$ and $V_{c2}$ are the primary and secondary side dc capacitor voltages, $s_1$ and $s_2$ are the primary and secondary full bridges converters' switching functions, $I_1$ and $I_2$ are primary side referred primary and secondary transformer winding current, $I_d$ is the input inductor dc current, $L_1$ and $L_2$ are the primary-side referred transformer winding self-inductances, $L_m$ is the transformers' magnetizing inductance, $R_1$ and $R_2$ are the transformer's winding resistances and $p = \frac{d}{dt}$, is a time-derivative operator

Again:

$$s_1 = s_{ap1} - s_{bp1}, \; s_2 = s_{ap2} - s_{bp2}, \; h = L_m^2 - L_1 L_2$$

$s_{ap1}$ and $s_{bp1}$ are the switching signals of the upper switches $\mathbf{S_{ap1}}$ and $\mathbf{S_{bp1}}$ of the primary bridge while $s_{ap2}$ and $s_{bp2}$ are the switching signals of the upper switches $\mathbf{S_{ap2}}$ and $\mathbf{S_{bp2}}$ of the secondary bridge.

### C. Steady-State Analysis.

To conduct the steady-state analysis of the system, the nonlinear time-periodic dynamic equations of the transformer winding currents are simplified, using the first harmonic approximations of the currents and the converter's switching functions. The switching functions which are quasi-square waves are expressed by their Fourier series as:

$$s_1 = \sum_{k=1,3,5,\ldots}^{\infty} \frac{4}{k\pi} \sin\left(k \frac{d_1}{2}\right) \sin(k\omega_s t) \quad (7)$$

$$s_2 = \sum_{k=1,3,5,\ldots}^{\infty} \frac{4}{k\pi} \sin\left(k \frac{d_2}{2}\right) \sin(k\omega_s t - k\delta) \quad (8)$$

Using the first harmonic approximation, (7) and (8) reduce to:

$$s_1 = real(S_{qd1} e^{j\omega_s t}), \quad s_2 = real(S_{qd2} e^{j\omega_s t}) \quad (9)$$

Where $S_{qd1}$ and $S_{qd2}$ are complex variables expressed by:

$$S_{qd1} = \frac{4}{\pi} \sin\left(\frac{d_1}{2}\right) \angle -\frac{\pi}{2}, \quad S_{qd2} = \frac{4}{\pi} \sin\left(\frac{d_2}{2}\right) \angle -\frac{\pi}{2} - \delta$$

The transformer winding currents which are responses due to switching functions in (9) are given by:

$$I_1 = real(I_{qd1} e^{j\omega_s t}), \quad I_2 = real(I_{qd2} e^{j\omega_s t}) \quad (10)$$

Where $I_{qd1}$ and $I_{qd2}$ are the primary and secondary complex-valued currents of the high frequency transformer winding.

Using (9) and (10) and the harmonic balance technique (HB) (3)-(6) are re-written as:

$$pV_{c1} = \frac{I_d}{C_1} - \frac{1}{2C_1} real(s_{qd1} I_{qd1}^*) \quad (11)$$

$$pI_{qd1} = \frac{L_1 R_1}{h} I_{qd1} + \frac{L_m R_2}{h} I_{qd2} - \frac{L_2}{h} s_{qd1} V_{c1} + \frac{L_m}{h} s_{qd2} V_{c2} - j\omega_s I_{qd1} \tag{12}$$

$$pI_{qd2} = \frac{L_m R_1}{h} I_{qd1} + \frac{L_1 R_2}{h} I_{qd2} - \frac{L_m}{h} s_{qd1} V_{c1} + \frac{L_1}{h} s_{qd2} V_{c2} - j\omega_s I_{qd2} \tag{13}$$

$$pV_{c2} = \frac{n}{2C_2} real(s_{qd2} I_{qd2}^*) - \frac{I_o}{C_2} \tag{14}$$

Where $f_{qd} = f_q + jf_d$ is the complex variable of the ac state variables.

To analyze these equations, they are split into real and imaginary parts. At steady state, the time-derivative of the state variables is equal to zero. Thus, the steady state equations of the system in Figure 1 are written as:

$$V_{c1} = V_1 \tag{15}$$

$$S_{q1} I_{q1} + S_{d1} I_{d1} = 2I_d \tag{16}$$

$$\frac{L_1 R_1}{h} I_{q1} + \frac{L_m R_2}{h} I_{q2} - \frac{L_2}{h} s_{q1} V_{c1} + \frac{L_m}{h} s_{q2} V_{c2} + \omega_s I_{d1} = 0 \tag{17}$$

$$\frac{L_1 R_1}{h} I_{d1} + \frac{L_m R_2}{h} I_{d2} - \frac{L_2}{h} s_{d1} V_{c1} + \frac{L_m}{h} s_{d2} V_{c2} - \omega_s I_{q1} = 0 \tag{18}$$

$$\frac{L_m R_1}{h} I_{q1} + \frac{L_1 R_2}{h} I_{q2} - \frac{L_m}{h} s_{q1} V_{c1} + \frac{L_1}{h} s_{q2} V_{c2} + \omega_s I_{d2} = 0 \tag{19}$$

$$\frac{L_m R_1}{h} I_{d1} + \frac{L_1 R_2}{h} I_{d2} - \frac{L_m}{h} s_{d1} V_{c1} + \frac{L_1}{h} s_{d2} V_{c2} - \omega_s I_{q2} = 0 \tag{20}$$

$$S_{q2} I_{q2} + S_{d2} I_{d2} = \frac{2I_o}{n} \tag{21}$$

(16) – (21) are solved by making it as an optimization problem. the steady-state equations are nonlinear and over-determined. First the control variables, $S_{q1}$, $S_{q2}$ and $S_{d2}$ are solved by minimizing the sum of the squares of peak of the fundamental harmonic of the transformer winding currents subject to the input and output dc current constraints while $S_{d1} = 0$

$$\min \left( |I_{qd1}|^2 + |I_{qd2}|^2 \right)$$

Subject to:

$$S_{q1} I_{q1} + S_{d1} I_{d1} = 2I_d$$

$$S_{q2} I_{q2} + S_{d2} I_{d2} = \frac{2I_o}{n}$$

Using the parameter values in Table 1, the optimization problem was solved numerically. Results obtained, depicted in Figure 3 show that the optimal values of the outer phase shift $\delta$ between the primary and secondary ac voltages is positive for forward active power flow and negative for backward active power flow. Again, for the forward active power flow (0 to 1kW), inner phase shift of the primary bridge is $180°$, while the secondary bridge assumes inner phase shift values between 0 and $180°$. The inner phase shift values for the primary bridge are between 0 and $180°$ and $180°$ for the secondary bridge for backward active power flow. Figure 4 shows the result for the optimal values of the primary-side-referred peak primary and secondary current of transformer windings. Figure 5 depicts the relationship between the input and output of the DAB.

TABLE I. DAB PARAMETERS

| Parameters | Values |
|---|---|
| Input voltage | $V_1 = 100\,V$ |
| Transformer winding self and mutual inductances | $L_1 = 4.2134\,mH$, $L_2' = 4.2158\,mH$, $L_m = 4.205\,mH$ |
| Input Inductance | $L_d = 10\,mH$ |
| Transformer, winding resistances | $R_1 = 0.45\Omega$, $R_2' = 0.45\Omega$ |
| Turns ratio | $n = 0.5$ |
| Desired output voltage | $V_{c2}^* = 200V$ |
| Desired Power | $-1kW\ to\ 1kW$ |
| Switching frequency | $f_s = 25kH$ |

### D. Zero Voltage Switching Analysis.

Based on the convention for the flow of current adopted for this paper, the conditions for ZVS are expressed in terms of the transformer winding currents at switching instants. The conditions are given as:

Half-Bridge 1 ($S_{ap1}, S_{an1}$): $I_1\left(\omega_s t = \pi - \frac{\alpha_1}{2}\right) > I_{1min}$

Half-Bridge 2 ($S_{bp1}, S_{bn1}$): $I_1\left(\omega_s t = \frac{\alpha_1}{2}\right) < -I_{1min}$

Half-Bridge 3 ($S_{ap2}, S_{an2}$): $I_2\left(\omega_s t = \delta - \frac{\alpha_2}{2}\right) > I_{2min}$

Half-Bridge 4 ($S_{bp2}, S_{bn2}$): $I_2\left(\omega_s t = \delta + \frac{\alpha_2}{2}\right) > I_{2min}$

Where $\alpha_1 = \pi - d_1, \alpha_2 = \pi - d_2$, and $I_{1min}$ and $I_{2min}$ are the minimum absolute current required to charge the junction device capacitance completely within the deadtime. Figure 6 shows the result of the ZVS analysis. The results show that the first half bridge on the primary side does not meet ZVS requirements, for the entire power range while, the second half bridge meets ZVS requirements from -1kw up to -100Watts. The first half bridge on the secondary side only meets the ZVS requirement for output power of 850 Watts to the maximum power whereas. the second half bridge meets the ZVS requirement throughout the entire power range except between -0.5kWatts and 0.25kW range.

### E. Stability Analysis.

Based on the steady-state analysis results obtained, the stability of the system is analyzed using the FHA model of the system for a constant voltage operation and a constant power operation. From the eigenvalues of the DAB system for the constant voltage operation, it is observed that the system is stable for the entire power range considered as all the eigenvalues of the system are on the left-hand side of the complex plane. The eigenvalues for three power points are shown in Table 2. Also, from the eigenvalues obtained for the constant power operation, the DAB system is observed to be stable for low output active power, but unstable for power demand close to the maximum value for the considered system. Eigenvalues for three power point is also shown in Table 3.

### III. CONTROLLAABILITY AND OBSERVABILITY AND ZERO DYNAMICS OF THE SYSTEM

#### A. Controllability and Observabilitiy Analysis.

For a nonlinear system given as:

$$\dot{x} = f(x) + g(x)u \qquad (22)$$

$$y = h(x) \qquad (23)$$

$$h(x) = [h_1(x) \quad h_2(x) \quad \ldots \quad h_j(x)],$$

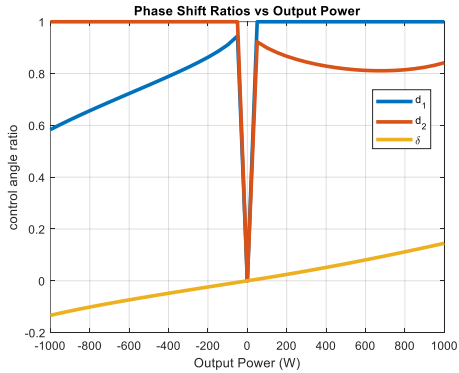

Fig. 3. Optimal solutions for normalized values of the outer phase shift $\delta$ (in yellow), duty cycle of the primary ac voltage, $d_1$ (in blue) and duty cycle of the secondary ac voltage, $d_2$ (in red)

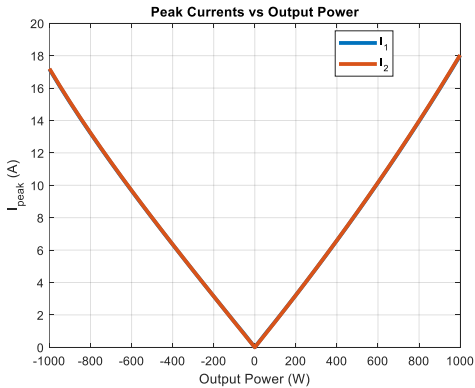

Fig. 4. Primary-side referred peak current values of first harmonic component of the primary winding current (in blue) and secondary winding current (in red) of the transformer

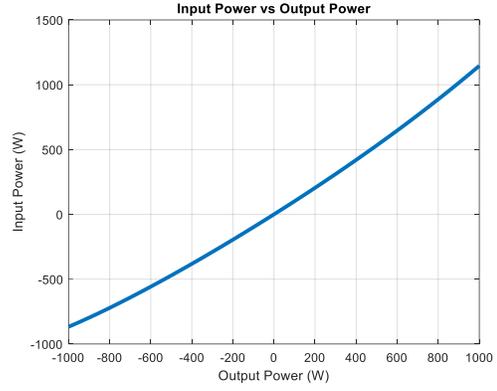

Fig. 5. Plot of input power vs output power

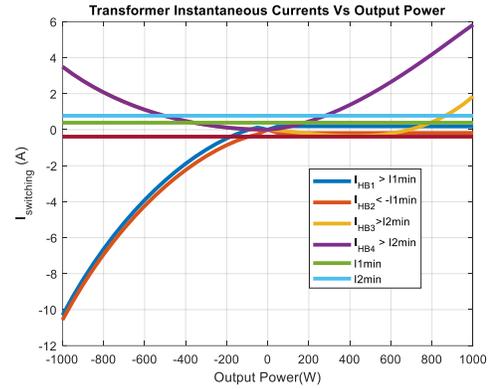

Fig. 6. Transformer winding currents at switching instants. Instantaneous current at $I_1\left(\omega_s t = \pi - \frac{\alpha_1}{2}\right)$ in blue, and instantaneous current at $I_1\left(\omega_s t = \frac{\alpha_1}{2}\right)$ in red

$$g(x) = [g_1(x) \quad g_2(x) \quad g_m(x)]$$

The controllability matrix is given by:

$$M = [g_1 \quad \ldots \quad g_m \quad ad_f^{k-1}g_1 \quad \ldots \quad ad_f^{k-1}g_m\ ]$$

Where $m$ is the number of control inputs, and $k$ is the number of outputs.

The rank of $M$ based on state space model equations ((2) – (6)) of the DAB system is 5 which is equal to the dimension of the system. This means the system has a full rank for the controllability matrix and the system is completely controllable. Again, for a nonlinear system, given a matrix $N$

$$N = [h_1 \quad L_f^{k-1}h_1 \quad \ldots \quad h_m \quad L_f^{k-1}h_m]$$

The observability matrix is given by:

$$J = \frac{dN}{dx} \ at\ x = x_o \qquad (24)$$

The system is locally weakly observable at $x = x_o$ if any k row combination in $J$ has full rank[7]. For the system considered, the maximum rank is three (3), which is less than the dimension of the system. This means the system is not fully observable.

TABLE II: CONSTANT VOLTAGE OPERATION EIGENVALUES

| S/N | 300W ($\times 10^5$) | 700W ($\times 10^5$) | 1000W ($\times 10^5$) |
|---|---|---|---|
| 1 | -0.4686+j1.5732 | -0.4686+j1.5732 | -0.4687+j1.5732 |
| 2 | -0.4686-j1.5732 | --0.4686-j1.5732 | -0.4687-j1.5732 |
| 3 | -0.0004+j0.0031 | -0.0004+j0.0031 | -0.0004+j0.0032 |
| 4 | --0.0004-j0.0031 | --0.0004-j0.0031 | --0.0004-j0.0032 |
| 5 | -0.0007 | -0.0007 | -0.0006 |
| 6 | -0.0005+j1.5708 | -0.0005+j1.5708 | -0.0005+j1.5708 |
| 7 | -0.0005-j1.5708 | -0.0005-j1.5708 | -0.0005-j1.5708 |

TABLE III: CONSTANT POWER LOAD EIGENVALUES

| S/N | 300W ($\times 10^5$) | 700W ($\times 10^5$) | 1000W ($\times 10^5$) |
|---|---|---|---|
| 1 | -0.4686+j1.5732 | -0.4686+j1.5732 | -0.4686+j1.5732 |
| 2 | -0.4686-j1.5732 | --0.4686-j1.5732 | -0.4686-j1.5732 |
| 3 | -0.0004+j0.0031 | -0.0004+j0.0031 | -0.0004+j0.0032 |
| 4 | --0.0004-j0.0031 | --0.0004-j0.0031 | -0.0004-j0.0032 |
| 5 | -0.0004 | 0 | 0.0003 |
| 6 | -0.0005+j1.5708 | -0.0005+j1.5708 | -0.0005+j1.5708 |
| 7 | -0.0005-j1.5708 | -0.0005-j1.5708 | -0.0005-j1.5708 |

*B. Zero Dynamics.*

Zero dynamics is the dynamics of a nonlinear system when the output states and reachable states of the system are constrained to zero [8]. It is the worst case of the internal dynamics of a nonlinear system. Thus, if the zero dynamics of a nonlinear system is stable, then the system can be stabilized by the input-output feedback control law. The relative degree of the system is determined by continuously differentiating the outputs of the system described by (2) – (6) until an input variable appears.

$$\dot{y}_1 = \dot{V}_{c1} = \frac{I_d}{C_1} - \frac{s_1 I_1}{C_1}, \quad r_1 = 1$$

$$\dot{y}_2 = \dot{V}_{c2} = \frac{n s_2 I_2}{C_2} - \frac{I_o}{C_2}, \quad r_2 = 1;$$

The relative degree of the system is thus $r = r_1 + r_2 = 2$

The number of zero dynamic states is: $z = k - r = 3$

Since $V_{c1}$ and $V_{c2}$ are the outputs, the zero dynamic states are $I_d$, $I_1$ and $I_2$. The zero dynamics equations are given by:

$$pI_d = \frac{V_1}{L_d} - \frac{V_{c1}}{L_d} - \frac{rI_d}{L_d} \tag{2}$$

$$pI_1 = \frac{L_1 R_1}{h} I_1 + \frac{L_m R_2}{h} I_2 - \frac{L_2}{h} s_1 V_{c1} + \frac{L_m}{h} s_2 V_{c2} \tag{4}$$

$$pI_2 = \frac{L_m R_1}{h} I_1 + \frac{L_1 R_2}{h} I_2 - \frac{L_m}{h} s_1 V_{c1} + \frac{L_1}{h} s_2 V_{c2} \tag{5}$$

Since (2) is decoupled from (4) and (5),

$$sI_d + \frac{r}{L_d} I_d = 0$$

$$s = -\frac{r}{L_d}$$

This means the input inductor current is stable.

At zero dynamics, $V_{c2} = 0$, $V_{c2} = 0$ and (4) and (5) reduce to:

$$pI_1 = \frac{L_1 R_1}{h} I_1 + \frac{L_m R_2}{h} I_2 \tag{25}$$

$$pI_2 = \frac{L_m R_1}{h} I_1 + \frac{L_1 R_2}{h} I_2 \tag{26}$$

If (25) and (26) are written as:

$$pI = AI \tag{27}$$

$$A = \begin{bmatrix} \frac{L_1 R_1}{h} & \frac{L_m R_2}{h} \\ \frac{L_m R_1}{h} & \frac{L_1 R_2}{h} \end{bmatrix} = \begin{bmatrix} a_{11} & a_{12} \\ a_{21} & a_{22} \end{bmatrix}$$

The characteristic equation is given by

$$|\lambda I - A| = 0 \tag{28}$$

$$\lambda^2 - (a_{11} + a_{22})\lambda + a_{11} a_{22} - a_{12} a_{21} = 0 \tag{29}$$

$$a_2 = 1, \; a_1 = -(a_{11} + a_{22}), \; a_o = a_{11} a_{22} - a_{12} a_{21}$$

$a_1$ and $a_2$ will always be positive, according to the Routh-Hurwitz criteria, necessary conditions for stability requires that: $a_o$ be always positive.

Sufficient conditions for stability states that $\Delta_1 = a_o > 0$

$$a_{11} a_{22} > a_{12} a_{21} \rightarrow L_1^2 > L_m^2$$

The primary winding self-inductance will always be greater that magnetizing inductance. This means regardless of the operating conditions of the system; the zero dynamics of the system will always be stable.

Again, using the dynamic equation based on the first complex variable form harmonic approximation of the actual model equations, the zero dynamics state remains the transformer winding current and the input inductor current. Hence the zero dynamics equations are given by:

$$pI_{qd1} = \frac{L_1 R_1}{h} I_{qd1} + \frac{L_m R_2}{h} I_{qd2} - j\omega_s I_{qd1} \tag{30}$$

$$pI_{qd2} = \frac{L_m R_1}{h} I_{qd1} + \frac{L_1 R_2}{h} I_{qd2} - j\omega_s I_{qd2} \tag{31}$$

(30) and (31) can also be written as:

$$pI_{qd} = AI_{qd} \tag{32}$$

$$A = \begin{bmatrix} \frac{R_1 L_1}{h} - j\omega_s & \frac{L_m R_2}{h} \\ \frac{R_1 L_m}{h} & \left(\frac{R_2 L_1}{h} - j\omega_s\right) \end{bmatrix}$$

It the characteristic equation is generally written as:

$$a_o \lambda^2 + (a_1 + jb_1)\lambda + a_2 + jb_2 \tag{33}$$

$$a_o = 1, \ a_1 = -\frac{L_1}{h}(R_1 + R_2), \ b_1 = 2\omega_s,$$

$$a_2 = \frac{L_1^2 R_1 R_2}{h^2} - \frac{L_m^2 R_1 R_2}{h^2} - \omega_s^2, \ b_2 = -\frac{\omega_s L_1}{h}(R_1 + R_2)$$

Routh-Hurwitz Stability criterion for the second order system is given as [9]:

$$\Delta_1 > 0; \quad \Delta_2 > 0$$

$$\Delta_1 = a_1, \ \Delta_2 = \det\left(\begin{bmatrix} a_1 & 0 & -b_2 \\ a_o & a_2 & -b_1 \\ 0 & b_2 & a_1 \end{bmatrix}\right)$$

$$\Delta_1 = a_1 > 0; \quad \Delta_2 = \frac{R_1 R_2 L_1^2}{h^4}(R_1 + R_2)^2 (L_1^2 - L_m^2) > 0$$

$$L_1^2 - L_m^2 > 0,$$

This stability condition is identical to the condition obtained based on the continuous-time system model. This further confirms that zero dynamics of the dual active bridge dc-dc converter is always stable regardless of the system operating conditions. The stability of the zero dynamics also explains the reason the neglect of the dynamics of the transformer current does not affect the design of a stable controller for the DAB using the input-output feedback method.

IV. CONCLUSION

In this article, the continuous time as well as the FHA model of the DAB are derived. Steady state analysis of the system is done by constructing it as an optimization problem where the sum of the squares of the transformer current peak was minimized. The stability analysis of the system is conducted, and results obtained shows that the system is always stable for a constant output voltage operation and unstable for a constant power load demand which is close to the DAB system's maximum output power. Finally, it is established that the zero dynamic of the DAB is the dynamics of the transformer currents. The zero dynamics analysis confirms that the DAB can be stabilized through the now popular input-output feedback control laws, neglecting the influence of the transformer currents, since they are stable, even though they are unobservable and may not be possibly estimated by an observer.